\documentclass[%
 twocolumn,
 superscriptaddress,
 amsmath,amssymb,
]{revtex4-2}

\usepackage[utf8]{inputenc}
\usepackage[OT1]{fontenc}
\usepackage[english]{babel}
\usepackage{soul}
\usepackage{siunitx}

\usepackage{graphicx}
  \graphicspath{{figures/}}
\usepackage{xcolor}

\usepackage{tabularx}

\usepackage{hyperref}

\begin{document}

\title{Ventilated noise-insulating metamaterials inspired by sonic black holes}

\author{Farid Bikmukhametov}
    \thanks{These authors contributed equally.}
    \affiliation{School of Physics and Engineering, ITMO University, Saint Petersburg 197101, Russia}

\author{Lana Glazko}
    \thanks{These authors contributed equally.}
    \affiliation{School of Physics and Engineering, ITMO University, Saint Petersburg 197101, Russia}

\author{Yaroslav Muravev}
    \thanks{These authors contributed equally.}
    \affiliation{School of Physics and Engineering, ITMO University, Saint Petersburg 197101, Russia}

\author{Dmitrii Pozdeev}
    \thanks{These authors contributed equally.}
    \affiliation{School of Physics and Engineering, ITMO University, Saint Petersburg 197101, Russia}
    
\author{Evgeni Vasiliev}
    \thanks{These authors contributed equally.}
    \affiliation{School of Physics and Engineering, ITMO University, Saint Petersburg 197101, Russia}

\author{Sergey Krasikov}
    \email{s.krasikov@metalab.ifmo.ru}
    \affiliation{School of Physics and Engineering, ITMO University, Saint Petersburg 197101, Russia}

\author{Mariia Krasikova}
    \affiliation{School of Physics and Engineering, ITMO University, Saint Petersburg 197101, Russia}

\date{\today}

\begin{abstract}
Acoustic black holes represent a special class of metastructures allowing efficient absorption based on the slow sound principle. The decrease of the wave speed is associated with the spatial variation of acoustic impedance, while the absorption properties are linked to thermoviscous losses induced by the local resonances of the structure. 
While most of the developments in the field of sonic black holes are dedicated to one-dimensional structures, the current study is concerned with their two-dimensional counterparts. It is shown that the change of the dimensionality results in the change of noise insulation mechanism, which relies on the opening of band-gaps rather then thermoviscous losses. The formation of band-gaps is associated with the strong coupling between the resonators constituting the considered structures. Numerically and experimentally it is shown than the structure is characterized by broad stop-bands in transmission spectra, while the air flow propagation is still allowed. In particular, a realistic application scenario is considered, in which the acoustic noise and the air flow are generated by a fan embedded into a ventilation duct.
The obtained results pave the way towards the development of next-level ventilated metamaterials for efficient noise control.
\end{abstract}

\maketitle

\section{Introduction}
The development of the low-frequency passive sound insulation systems still represent an unsolved challenge. Recent studies in the field of acoustic metamaterials~\cite{cummer2016controlling, assouar2018acoustic,liao2021acoustic} open a rout towards the development of low-weight sub-wavelength structures allowing not only efficient noise suppression~\cite{gao2022acoustic}, but also air ventilation~\cite{kumar2020recent, ang2023systematic}. A special class of metastructures is represented by the so-called acoustic black holes (ABH) in which the phase velocity reduces until zero due to spatial variation of acoustic impedance~\cite{krylov2014acoustic,pelat2020acoustic}. In this sense, the incoming wave is trapped inside the structure, in analogy with the corresponding cosmological objects. 
Successfully implemented for suppression of vibrations~\cite{zhao2019acoustic}, the concept of ABH was successfully translated for airborne sound, where the structures typically represent a set of plates characterized by the gradually changing sizes and distances between them~\cite{mironov2002onedimensional,ouahabi2015experimental, ouahabi2015investigation,mironov2020onedimensional}.
The absorption mechanism of ABH in this case relies on resonant states inducing the increase of thermoviscous losses~\cite{mousavi2022how,cervenka2022role,umnova2023multiple}. Though, in realistic systems the fulfillment of critical coupling condition is rather difficult to achieve and additional porous inserts might be required~\cite{hruska2024complex}. 
In order to distinguish between two different types of structures, ABHs for sound propagating fluids are usually referred to as sonic black holes (SBH), while the term vibrational black holes (VBH) is utilized for bending waves~\cite{mironov2020onedimensional}.

\begin{figure*}[htbp!]
    \centering
    \includegraphics[width=0.9\linewidth]{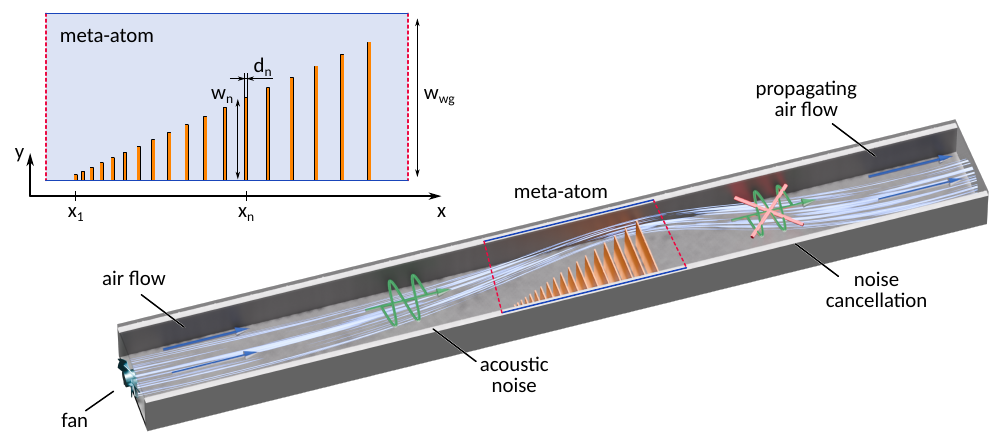}
    \caption{\textbf{Artistic illustration of the considered system.} The meta-atom consists of a set of plates each characterized by the coordinate $x_n$, the width $w_n$ and the thickness $d_n$. Placed in a waveguide, such a structure allows to block the propagation of acoustic noise while the air flow can propagate without a significant reduction of its speed.}
    \label{fig:system}
\end{figure*}

Typically, SBH are utilized as absorbing termination ends of circular waveguides, utilizing the ring-based geometry developed in Ref.~\cite{mironov2002onedimensional}, or ventilated duct silencers~\cite{mi2021wave, mi2022broadband, bravo2023broadband}. More complicated geometries may also include arrays of discs~\cite{mironov2020onedimensional} or combination of rings with other structures, such as lattices~\cite{chua2023novel}.
The circular geometry implies that the structure effectively is one-dimensional. The increase of the dimensionality might be associated with the periodicity of the structure in the direction perpindicular to the direction of the incident wave propagation. Wile periodic arranges of VBH and SBH were suggested for suppression of vibrations~\cite{semperlotti2015acoustic,deng2022metamaterial}, two-dimensional versions of SBH are still under-represented (with the rare exceptions like in Ref.~\cite{hruska2024complex}).
However, it might be speculated that the change of dimensionality might might allow to improve performance of the noise-insulating systems via the use of band-gaps.

Hence, the work is dedicated to the development of ventilated meta-atoms for efficient noise mitigation in rectangular ducts. Considered meta-atoms represent a set of rectangular plates forming a set of strongly coupled resonators. Noise-suppression properties of the meta-atoms are described in terms of band-structures of the associated infinitely periodic 2D systems. It is shown that wide stop-bands in the transmission spectra originate from the coupling between the resonances formed by the plates. As the consequence, the properties of the considered meta-atoms are not directly defined by the geometric profile of the structure. The obtained experimental results indicate that the broad stop-band covers nearly the whole range from $1000$ to $2850$~Hz. At the same time, the structure is ventilated and the meta-atom reduces the air flow speed by only $25$\%. These findings might be useful for the further development of metastructures for passive noise insulation in ventilated ducts.

\section{Materials and Methods}
\subsection{System description}
Considered meta-atoms consist of rectangular plates having different widths and arranged at unequal distances with respect to each other (see Fig.~\ref{fig:system}). Similar structures frequently investigated as sonic black holes are usually characterized by axial symmetry meaning that effectively they are one-dimensional, contrary to the considered case. When the meta-atom is placed inside a rectangular duct with sound hard walls, it can be considered as an effectively two-dimensional structure, which is periodic along the directions perpendicular to the direction of the incident wave propagation. The approximation is valid as long as the field distribution is characterized by the mirror symmetry with respect to the direction of the wave propagation, i.e. $x$-axis. Similarly, the system is effectively two-dimensional until the field distribution along the $z$-axis is uniform. Mirror symmetry also implies that instead of the whole structure, only its half can be utilized without the loss of generality. The verification of this statement, as well as the detailed description of the design procedure are given in Supplementary Information.

\begin{figure*}[htbp!]
    \centering
    \includegraphics[width=0.9\linewidth]{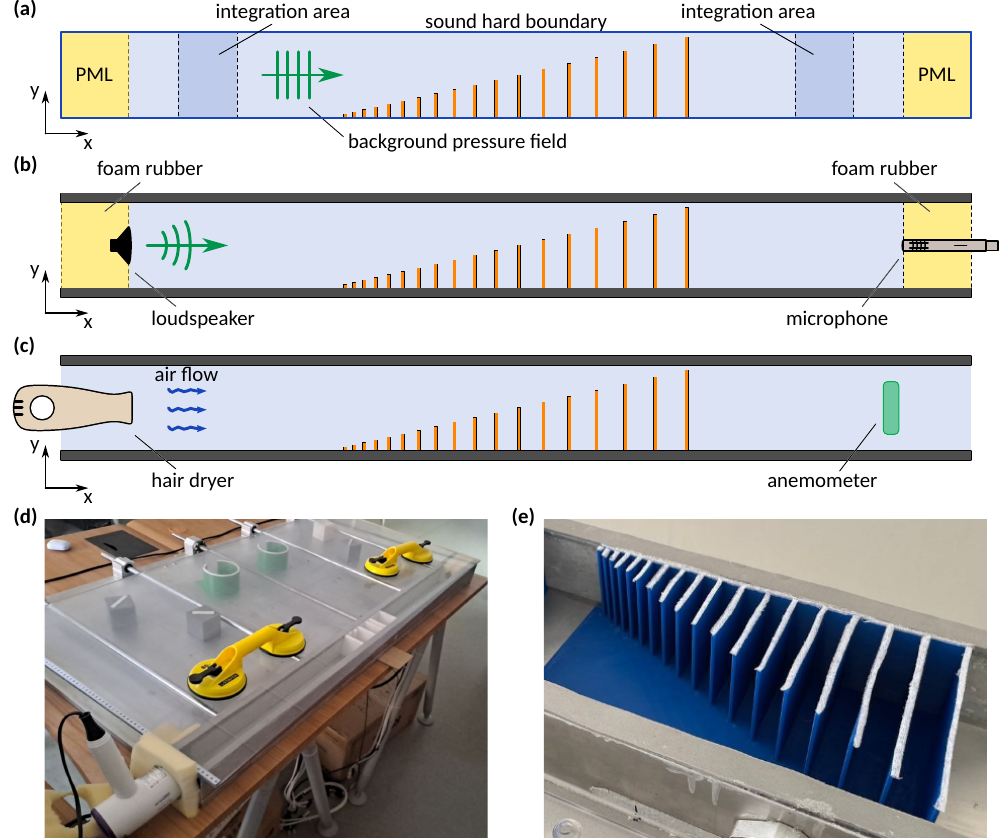}
    \caption{\textbf{Illustration of the utilized methods.} (a) Schematics of the numerical model. Schematics of the experimental setup for (b) transmission measurements and (c) measurements of the air flow speed. (d) Photo of the experimental setup. (e) Photo of a sample.}
    \label{fig:methods}
\end{figure*}

\subsection{Numerical Calculations}
Numerical calculations are provided in COMSOL Multiphysics. Transmission spectra calculations presented in the main text are obtained using the ``thermoviscous acoustics, frequency domain'' physics. The incident wave is implemented as the background pressure field with the amplitude $p_0 = 1$~Pa. Both ends of the waveguide are supplemented by the perfectly matched layers, imitating absorbing termination ends of the experimental setup. All walls of the resonators are considered to be the sound hard ones. Still, viscous boundary layers are taken into account, such that the thickness of the boundary layers are considered to be 
\begin{equation}
    d_{\mathrm{visc}} = d_{\mathrm{visc}, 0} \sqrt{f_0/f},
\end{equation}
where $d_{\mathrm{visc}, 0} = 0.22$ mm is the boundary layer thickness at the frequency $f_0 = 100$ Hz. Correspondingly, the mesh is supplemented by the ``boundary layers'' feature.

Transmission coefficients given in decibels are obtained as
\begin{equation}
    T = 20 \log_{10} P_t/P_b,
\end{equation}
where $P_t$ and $P_b$ are the amplitudes of the total and background pressure fields, respectively, calculated via the integration of the pressure over the area behind the structure [see Fig.~\ref{fig:methods}(a)]:
\begin{equation}
    P_{t,b} = \frac{1}{A} \int\limits_{A} |p_{t,b}|^2 dA.
\end{equation}
Note that transmission spectra are calculated for the halved unit cell, contrary to the case of the eigenmodes analysis. 

\subsection{Optimization}
The optimization procedure is based on a simple genetic algorithm, similar to the one implemented in Ref.~\cite{krasikova2024broadband}. In particular, the goal of the algorithm is to find geometric parameters of the structure for which the value of the cost function $\mathcal{C}$ is minimized. The considered cost function represents a linear combination of the transmission coefficient averaged over the whole spectra $\langle T \rangle$, the transmission coefficient averaged over the specified part of the spectra $\langle T \rangle_{[f_1,f_2]}$, and its standard deviation $\sigma_{[f_1,f_2]}$:
$$
\mathcal{C} = a_1 \langle T \rangle + a_2 \left( \langle T \rangle_{[f_1,f_2]} + \sigma_{[f_1,f_2]} \right).
$$
All calculations are performed for the spectral range $100$ - $3000$ Hz, while $\langle T \rangle_{[f_1,f_2]}$ is defined within the target range $300$ - $2000$ Hz. Such a combination of parameters allows to ensure that the stop-band occurs in the desired range, and the transmission spectra in this case is close to a flat one. At the same time, the consideration of $\langle T \rangle$ allows to ensure that there are no undesired resonances outside of the target range, which might be crucial for practical applications. The procedure is performed for different values of the coefficients $a_{1,2}$. Namely, the structure discussed in the main text is obtained with $a_1 = 0.1$ and $a_2 = 0.9$. 

For the optimization procedure, the size of the structure is limited by $10$ plates. Each plate is characterized by the width, which can take a value between $0$ and $55$ mm (the width of the waveguide is fixed at $60$ mm). Similarly, the thickness of each plate and the distance between the plates can take values from the range from $1$ to $10$ mm. 

The population size is considered to be $20$, such that the initial population is generated randomly. The selection procedure is implemented using the roulette wheel approach with the selection pressure $0.1$. For the crossover operation, single point, double point, or uniform crossover operators are utilized, such that the particular operator can be selected with the probability $0.2$, $0.2$, and $0.6$, respectively. The mutation is based on the normal distribution with the standard deviation $0.1$, and the mutation rate is fixed at $0.1$.

\subsection{Measurements}
Experimental measurements are performed in the rectangular waveguide with the height $60$~mm and the adjustable width, while the length of the waveguide is $1350$~mm. The walls and the bottom of the waveguide are made of $15$~mm thick aluminium and the lid is made of plexiglass with the thickness $6$~mm. Sound waves are generated by the loudspeaker (Visaton BF 45/4) located at one end of the waveguide, and the recording of the transmitted field is done with the shotgun microphone (RODE NTG4) placed at the other end of the waveguide. To avoid reflections at the end of the waveguide, both the loudspeaker and the microphone are embedded into the porous inserts made of melamine foam. The length of the inserts is $150$~mm. The schematics of the setup and its photo are shown in Figs.~\ref{fig:methods}(b) and~\ref{fig:methods}(d).

Generation and recording of the signals is controlled via the custom software based on the sounddevice python module~\cite{sounddevice}. For that both the loudspeaker and the microphone are connected to the USB audio interface Roland Rubix22, such that the loudspeaker is connected via the amplifier based on the Yamaha YDA138-E microchip. Generated signals are the chirped signals with the duration $10$~s and the sampling frequency $44100$~Hz. For each measurement the signal was generated and recorded $6$ times and then averaged in the frequency domain. The transmission spectra in this case is defined as
\begin{equation}
    T = 20 \log_{10} p_{\mathrm{tr}}/p_{\mathrm{ref}},
\end{equation}
where $p_{\mathrm{tr}}$ is a pressure amplitude of the transmitted wave and $p_{\mathrm{ref}}$ is the reference pressure amplitude of the wave in the absence of the structure. The same procedure is performed for the measurements with the noise generated by a fan, i.e. the loudspeaker is substituted by the hair dryer (STINGRay ST-HD801A). The measured signal-to-noise ratio for both the loudspeaker and the hair dryer are presented in the Supplementary Information.

Ventilation measurements are performed in the same waveguide. The airflow is generated by the hair dryer, the same as for the noise measurements [see Fig.~\ref{fig:methods}(c)]. The speed of the flow is measured with the anemometer (ADA AeroTemp 30) located at the distance $55$~cm from the structure, such that the distance between the hairdryer and the structure is also about $55$~cm. 

The samples are manufactured by 3D printing using a PLA filament. To avoid gaps between the lid and the structure, the tops of the plates were supplemented by a thin layers of a porous material [see Fig.~\ref{fig:methods}(e)].

\section{Results}
\subsection{``Spruce'' Meta-Atom}

\begin{figure*}[htbp!]
    \centering
    \includegraphics[width=0.9\linewidth]{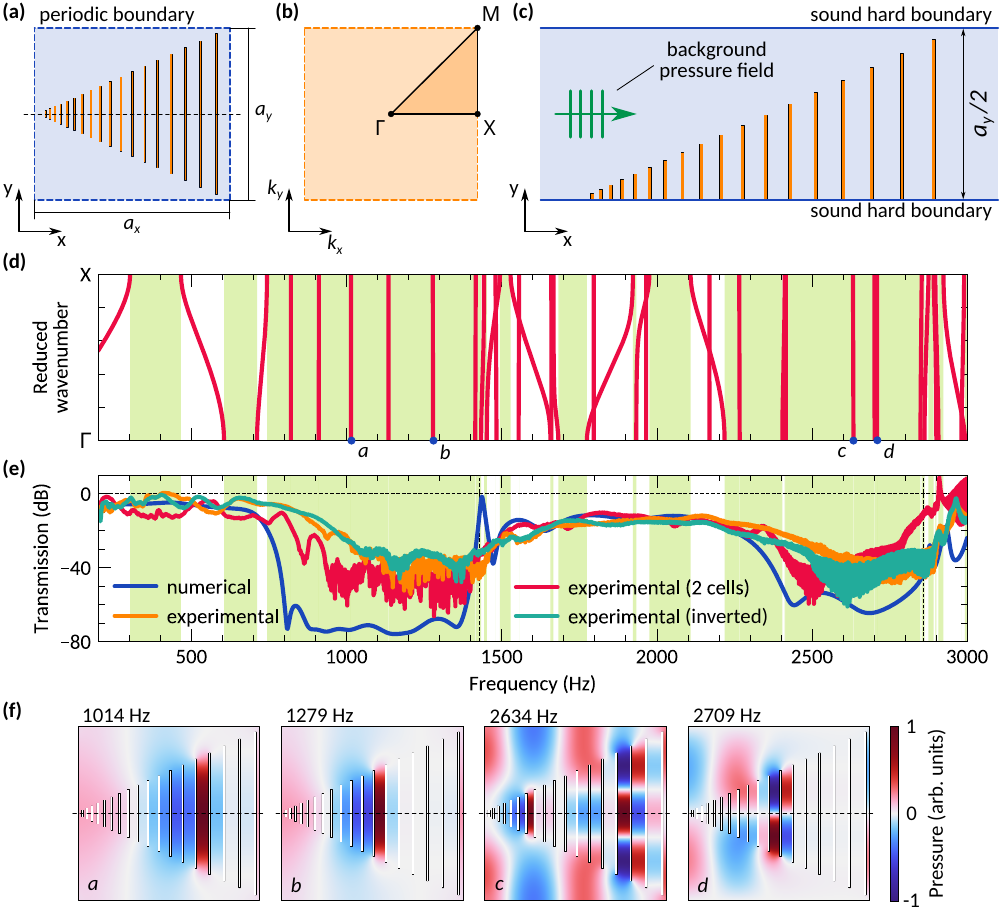}
    \caption{\textbf{``Spruce'' meta-atom.} (a) The considered unit cell and (b) the corresponding Brillouin zone. The unit cell is characterized by the width $a_x$ and the height $a_y$. Horizontal dashed line indicates the reflection symmetry axis. (c) Schematic illustration of the finite-size system consisting of the halved meta-atom placed into a waveguide with sound hard boundaries. The width of the waveguide is $w_{wg} = a_y/2$. (d) Band structure of the infinitely periodic system. (e) Transmission spectra of the finite-size structure having a thickness of one (blue and orange curves) and two (red curve) meta-atoms. The emerald line corresponds to the inverted structure, such that the largest plate is located closer to the loudspeaker and the smallest one -- closer to the microphone (i.e. the meta-atom is flipped with respect to the $y$-axis).
    Shaded green areas indicate band gaps of the associated infinitely periodic system. Vertical dashed lines correspond to the frequencies at which the wavelength is equal to $w_{wg}/2$ and $w_{wg}$. (f) Field distributions corresponding to the eigenmodes (at $\Gamma$ point) labelled as \textit{a}--\textit{d} on the band diagram.}
    \label{fig:spruce}
\end{figure*}

The considerations start from the structure in which parameters of the plates change gradually, such that the coordinates of the plates along the $x$-axis are defined as 
\begin{equation}
    x_n = 5(n+1) + n(n-1)/2,
\end{equation}
where $n$ is the number of a plate, and the corresponding semi-widths are 
\begin{equation}
    w_n = 115 x_n/255.
\end{equation}
The thickness of all plates is fixed at $d = 2$~mm. In the consequent text, such meta-atom will be refereed to as a ``spruce'', just due to associations with some schematic pictures of the corresponding tree.
One-dimensional structures with the similar parameters were considered in Ref.~\cite{mousavi2022how}, where it was shown that attenuation inside the structure is associated with the resonances of the cavities formed by the plates. As in the considered two-dimensional case the structure is periodic, it is reasonable to start from the consideration of its band structure. The corresponding unit cell [see Figs.~\ref{fig:spruce}(a) and~\ref{fig:spruce}(b)] is characterized by the width $a_x$ and the height $a_y$, which are considered to be $250$ and $240$~mm, respectively. According to the band structure shown in Fig.~\ref{fig:spruce}(d), the system is characterized by a set of band-gaps, which are practically merged in the regions $750$ -- $1450$~Hz and $2200$ -- $2800$~Hz, as the modes within this ranges are flat. The field distributions shown in Fig.~\ref{fig:spruce}(f) indicate that these modes correspond to the resonances in the cavities formed by the plates. Practically, the meta-atom can be considered as a set of resonators strongly coupled to each other. This coupling is the origin of the flat-band formation and the opening of band-gaps as discussed in the Supplementary Material. 

It should be expected, that when the equivalent finite-size structure is considered, the transmission spectra should be characterized by the stop-bands corresponding to the band-gaps of the infinite structure. Indeed, as shown in Fig.~\ref{fig:spruce}(e), the experimentally measured transmission drops below $-40$~dB in the corresponding regions $900$ -- $1450$~Hz and $2450$ -- $2850$~Hz. At the same time, not all band-gaps are manifested as the stop-bands. The reason for that is the small thickness of the structure, which is just a single unit cell. When the thickness increases, more stop-bands can be observed (see Supplementary Information). In particular, experimentally measured transmission spectra of a system having a thickness of two meta-atoms [see Fig.~\ref{fig:spruce}(e)] is characterized by a small dip within the range $300$ -- $450$~Hz, where a band-gap is located. In addition, the two other stop-bands are a bit wider and deeper than for the case of the structures consisting of a single meta-atom. It should be also noted that the experimental results do not quite match the numerical ones, which is associated with the fact that in the real system the finite height of the samples and the presence of the lid and bottom of the waveguide result in the occurrence of additional thermoviscous losses, which are not taken into account in the calculations. In addition, geometric parameters of the samples might differ from the ones used in the numerical calculations due to the limited accuracy of the utilized 3D printer. Still, the qualitative behavior of the measured and calculated spectra is the same. In addition, an inverted structure is considered, such that the meta-atom is flipped inside the waveguide, i.e. the largest plate is located closer to loudspeaker and the the smallest one -- closer to the microphone. The resulting transmission spectra [see Fig.~\ref{fig:spruce}(e)] is practically the same as for the initial orientation of the meta-atom, additionally indicating that the properties of the structure are determined by the resonances, rather than the gradual change of the admittance. Hence, the structure works as a reflector, not an absorber.

\begin{figure}[htbp!]
    \centering
    \includegraphics[width=0.9\linewidth]{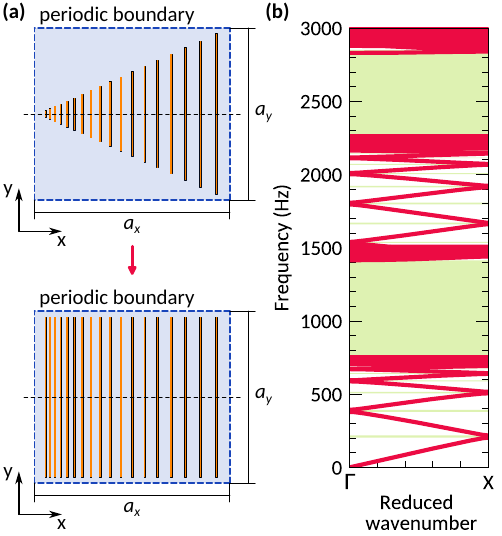}
    \caption{\textbf{Reciprocal SBH}. (a) The considered unit cell characterized by the width $a_x$ and the height $a_y$. Horizontal dashed line indicates the reflection symmetry axis. The initial unit cell of the ``spruce'' meta-atom is deformed in such a way that all plates have the same width. (d) The corresponding band structure of the infinitely periodic system. Green shaded areas indicate the $\Gamma$-$X$ band-gaps.
    }
    \label{fig:spruce_tilt}
\end{figure}

Still, an effect similar to the acoustic black hole can be observed for the case when the size width of the plates is equalized [see Fig.~\ref{fig:spruce_tilt}(a)]. In this case the band structure is characterized by two band-gaps, but the most interesting observation is the fact that the tilt of the low-frequency eigenmodes decreases with the increase of frequency until the modes become flat [see Fig.~\ref{fig:spruce_tilt}(b)]. These flat bands correspond to the lower boundary of the first band-gap. The same situation repeats for the modes between the first and the second band-gaps. As the inclination of the modes indicate the group velocity, defined as $v_g = \partial \omega/\partial k$, it can be stated that the group velocity decreases with the increase of the frequency until reaching zero, following by the opening of a band-gap. This is a direct analogy of a conventional ABH effect in which the phase velocity decreases with the propagation coordinate. In this sense, the considered structure can be regarded as a \textit{reciprocal ABH}. The ``spruce'' meta-atom in this case is practically the deformed version of such a structure allowing the formation of larger number of band-gaps, which is the aim of the work. While the further exploration of the effect lies beyond the scope of the work, some additional details are provided in Supplementary Information.

\begin{figure*}[htbp!]
    \centering
    \includegraphics[width=0.9\linewidth]{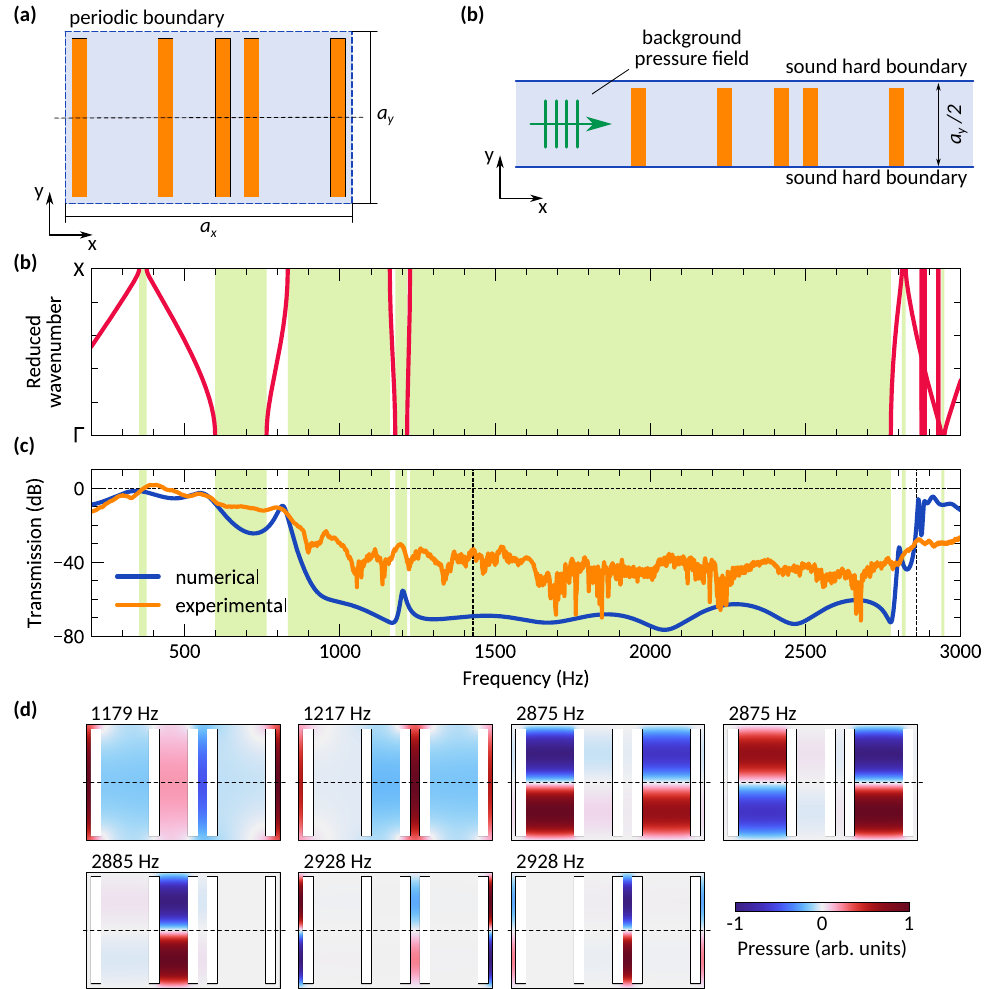}
    \caption{\textbf{Optimized meta-atom.} (a) The considered unit cell characterized by the width $a_x$ and the height $a_y$. Horizontal dashed line indicates the reflection symmetry axis. (b) The corresponding unit cell in the reciprocal space.  (c) Schematic illustration of the finite-size system consisting of the halved meta-atom placed into the waveguide with sound hard boundaries. The width of the waveguide is $w_{wg} = a_y/2$. (d) Band structure of the infinitely periodic system. (e) Transmission spectra of the finite-size structure. Shaded green areas indicate band gaps of the associated infinitely periodic system. Vertical dashed lines correspond to the frequencies at which the wavelength is equal to $w_{wg}/2$ and $w_{wg}$. (f) Field distributions corresponding to the eigenmodes (at $\Gamma$ point).
    }
    \label{fig:optimized}
\end{figure*}

\subsection{Optimized meta-atom}
While the ``spruce'' meta-atom demonstrate noticeable reduction of the transmission coefficient in the expected spectral regions, it might be argued that the geometric parameters of the structure are not optimized and the performance might be improved. The particular aim is to merge the stop-bands and simultaneously decrease the size of the structure. Following the analysis of coupled resonators and eigenmodes of periodic systems (see Supplementary Information), it might be expected that the optimized structure should consist of several nearly equivalent cavities characterized by a large width, as it allows to increase the width of band-gaps and shift them towards the low frequencies.
Using the optimization procedure described in Materials and Methods, the meta-atom shown in Fig.~\ref{fig:optimized}(a) is obtained. It consists of just $5$ plates forming four cavities inside a unit cell and two additional cavities formed by the plates from the neighboring unit cells (along the $x$-axis). The geometric profile in this case is flat, meaning that all plates have the same width, contrary to the ``spruce'' meta-atom and typical geometries of SBH. There is also no gradual change of the cavity sizes, as it not an important requirement for the adjustment of the coupling between the resonators. The band structure of the infinitely periodic system consisting of such meta-atoms is characterized by broad band-gaps covering almost the whole range from $600$ to $2800$ Hz [see Fig.~\ref{fig:optimized}(c)]. Again, the modes correspond to the resonances in-between the cavities [see Fig.~\ref{fig:optimized}(e)], which can be characterized by both symmetric and anti-symmetric distributions. Here it also should be mentioned, that resonators also interact with their neighbors along the $y$-axis, which also affects the results (see Supplementary Information). Notably, such effect is unavailable in one-dimensional SBHs, which make a distinguish between these two types of systems. 
The corresponding transmission spectra of the finite-size structure is characterized by the broad stop-band spectrally coinciding with the band-gaps of the infinitely periodic structure [see Figs.~\ref{fig:optimized}(b) and ~\ref{fig:optimized}(d)]. Again, there qualitative differences between the numerical and experimental results caused by the limitations of the experimental setup, but still the transmission within the stop-band is mostly below $-40$ dB, while the stop-band itself is much larger than for the case of the ``spruce'' meta-atom.

\begin{figure*}[htbp!]
    \centering
    \includegraphics[width=0.9\linewidth]{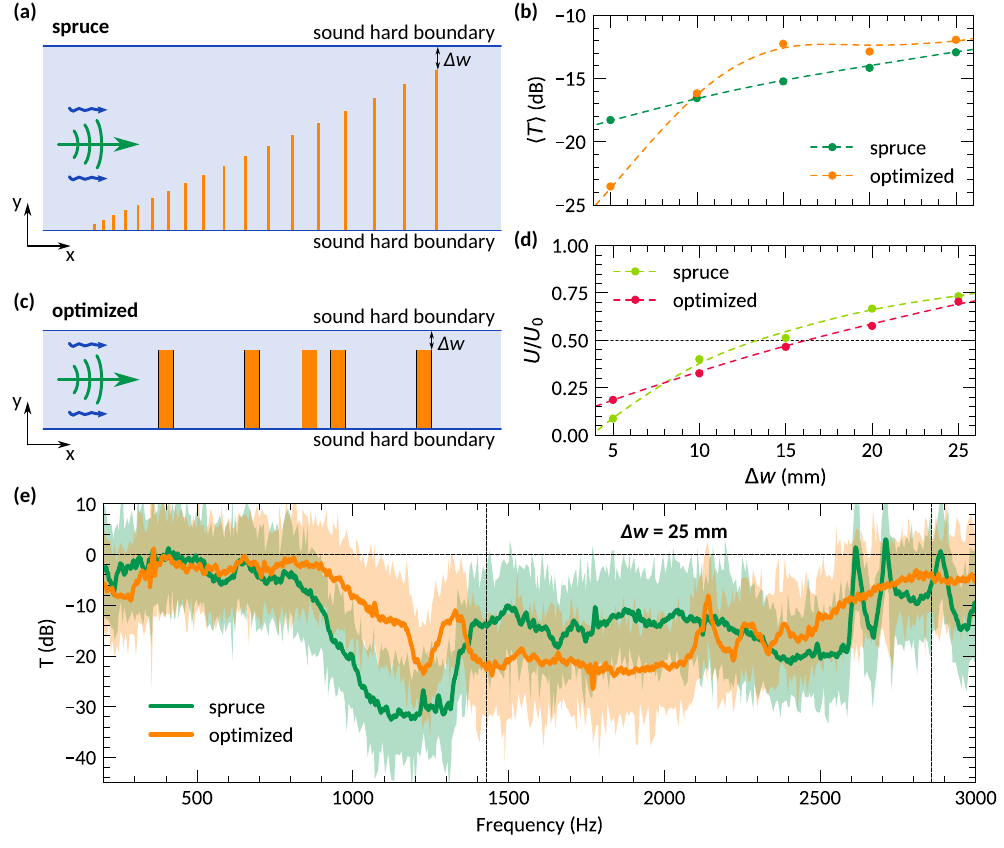}
    \caption{\textbf{Meta-atoms in a ventilated duct.} (a), (c) Schematic illustration of the considered meta-atoms placed inside a rectangular waveguide. Incident acoustic field and the air flow are generated by the hair dryer. (b) The transmission coefficient averaged in the range $300$ -- $2000$ Hz and (d) ratio of the flow speed with and without the structure as functions of the distance between the meta-atoms and the waveguide boundary. (e) Transmission spectra of the ``spruce'' and optimized meta-atoms for the case when the gap is $\Delta w = 25$~mm. Shaded areas indicate the actual noisy spectra and the solid lines correspond to the same data smoothed with the Savitzky–Golay filter. Vertical dashed lines correspond to the frequencies at which the wavelength is equal to $w_{wg}/2$ and $w_{wg}$.}
    \label{fig:ventilation}
\end{figure*}

\subsection{Ventilation}
Geometry of the structure implies the presence of the gaps between the meta-atom and one of the waveguide walls. Therefore, it might be expected that an airflow propagating through the waveguide will not be fully blocked by the meta-atom. In particular, the specific case of the noise control in the ventilation ducts in which both acoustic noise and air flow propagate simultaneously. As discussed in the Materials and Methods, to test the structure under such conditions, a fan is integrated into one the waveguide termination ends. Obviously, both transmission and ventilation depend on the size of the gap between the meta-atom and the wall of the waveguide [see Figs.~\ref{fig:ventilation}(a) and ~\ref{fig:ventilation}(c)]. It might be expected that the increase of the gap will result in the corresponding increase of the air flow speed behind the structure and, at the same time, in the increase of the transmission coefficient.

Indeed, as shown in Fig.~\ref{fig:ventilation}(b), the averaged transmission coefficient increases  when the gap becomes larger. For the case of the ``spruce'' meta-atom, the averaged transmission coefficient changes almost linearly, such that the difference between the lowest and the largest value is just $-5$~dB. For the optimized meta-atom the situation is a bit more complicated, and the dependence of the transmission coefficient on the gap size resembles a logarithmic function. The change in the values is rather drastic, about $15$~dB. Notably, the optimized structure demonstrate much better noise-insulating performance at low values of $\Delta w$, than the ``spruce'', which was actually the aim of the optimization procedure. When the gap size increases, and becomes more than $10$ mm, the ``spruce'' structure demonstrates better performance, indicating the fact that the optimization procedure should be actually performed for each value of $\Delta w$.

The ventilation properties of both meta-atoms is quite similar to each other [see Fig.~\ref{fig:ventilation}(d)]. Expectantly, the ratio of the air flow speed $U$ and the reference flow speed  $U_0$ (i.e. the speed without the structure) increases with the increase of the gap between a structure and the waveguide. The performance of the ``spruce'' meta-atom is slightly better, than for the optimized one, which might be associated with its geometry implying the gradual decrease of the effective channel width. In any case, the ventilation above $50$\% is achieved when the value of $\Delta w$ exceeds $15$~mm. For $\Delta w = 25$~mm the ratio $U/U_0$ is nearly $0.75$, meaning that only $25$\% of the air flow is blocked. The transmission spectra in this case is still characterized by stop-bands [see Fig.~\ref{fig:ventilation}(d)] occurring within the range $1000$--$2800$ Hz. Contrary to the previous measurements with the loudspeaker, the transmission spectra for the case of the hair dryer is noisy, but distinguishable gaps are still clearly visible. Therefore, a compromise between noise-insulation and ventilation properties can be found for the realistic application scenario.


\section{Conclusion}
To conclude, the work is dedicated to the investigation of 2D structures having the geometries similar to those of 1D sonic black holes. It is shown that the change of the dimensionality practically ruins the black hole effect and the decrease in transmission is associated with the opening of band-gaps, such that the structure works as a reflector rather than absorber. The formation of band-gaps is associated with the coupling between the resonators constituting the structure. As a consequence, the geometric profile in the 2D case is not the key factor, and only the size of the cavities matters. Still, an analog of a sound black hole effect is numerically observed, such that the group velocity gradually decreases with frequency until reaching zero, though this effect is not utilized in the work. As the main result, it is shown that fro the realistic scenario of a ventilation duct in which both the airflow and acoustic noise propagate simultaneously, it is possible to find a compromise between ventilation properties and noise insulation.
In particular, it is shown that the reduction of the air flow speed might be about $25$\% while the broad stop-band is observed within the spectral range $1000$--$2850$~Hz.
These results might be useful for further development of ventilated acoustic metamaterials and noise insulating systems for passive noise insulation, especially in ventilation ducts.

\section*{Author Contributions}
FB, LG, YM, DP, and EV contributed equally to this work. 
LG and SK provided the numerical calculations. FB, LG, YM, DP, EV and MK performed the experimental measurements.
SK proposed the idea and supervised the project. MK acquired the funding and co-supervised the project.

\section*{Declaration of Competing Interest}
The authors declare that they have no known competing financial interests or personal relationships that could have appeared to influence the work reported in this paper.

\section*{Data availability}
The data that support the findings of this study are available from the corresponding author upon reasonable request.

\section*{Acknowledgements}
The authors thank Aleksandra Pavliuk for the help with the experimental measurements and Mikhail Kuzmin for fabrication of the samples. The authors also thank Anton Melnikov for fruitful discussions and useful comments. 
This work is supported by the Russian Science Foundation (project 24-21-00275). 


\bibliographystyle{elsarticle-num}
\bibliography{bibliography}

\end{document}


\title{Supplementary Information for
\\
Ventilated noise-insulating metamaterials inspired by duct acoustic black holes}

\author{Farid Bikmukhametov}
    \thanks{These authors contributed equally.}
    \affiliation{School of Physics and Engineering, ITMO University, Saint Petersburg 197101, Russia}

\author{Lana Glazko}
    \thanks{These authors contributed equally.}
    \affiliation{School of Physics and Engineering, ITMO University, Saint Petersburg 197101, Russia}

\author{Yaroslav Muravev}
    \thanks{These authors contributed equally.}
    \affiliation{School of Physics and Engineering, ITMO University, Saint Petersburg 197101, Russia}

\author{Dmitrii Pozdeev}
    \thanks{These authors contributed equally.}
    \affiliation{School of Physics and Engineering, ITMO University, Saint Petersburg 197101, Russia}
    
\author{Evgeni Vasiliev}
    \thanks{These authors contributed equally.}
    \affiliation{School of Physics and Engineering, ITMO University, Saint Petersburg 197101, Russia}

\author{Sergey Krasikov}
    \email{s.krasikov@metalab.ifmo.ru}
    \affiliation{School of Physics and Engineering, ITMO University, Saint Petersburg 197101, Russia}

\author{Mariia Krasikova}
    \affiliation{School of Physics and Engineering, ITMO University, Saint Petersburg 197101, Russia}

\date{\today}

\maketitle

\tableofcontents

\clearpage

\section{Methods}
\begin{figure}[htbp!]
    \centering
    \includegraphics[width=0.9\linewidth]{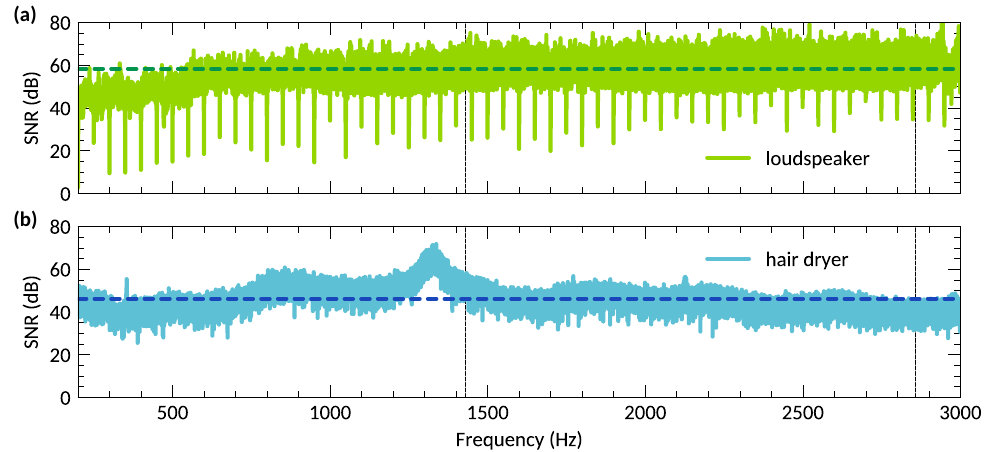}
    \caption{\textbf{Signal to noise ratio (SNR) of the setup.} Measured SNR for the case of (a) lodspeaker and (b) hair dryer. Horizontal dashed lines indicate the spectrally averaged SNR within the range $200$ -- $3000$~Hz. In both cases the width of the waveguide is $w_{\mathrm{wg}} = 120$~mm. Vertical dashed lines correspond to the frequencies at which the wavelength is equal to $w_{wg}/2$ and $w_{wg}$.}
    \label{fig:snr}
\end{figure}

The signal-to-noise ratio (SNR) of the setup is measured for both cases of sound sources, i.e. the loudspeaker and the hairdryer, when the width of the waveguide is $120$ mm. As shown in Fig.~\ref{fig:snr}(a), the SNR spectra is characterized by peaks occurring at each $50$ Hz, which are presumably noises related to the power supply circuit of the loudspeaker. These peaks are eliminated in the transmission measurements since the recorded signal is divided by the reference. The spectrally averaged SNR within the range from $200$ to $3000$ Hz is about $58$ dB.
For the hair dryer the SNR spectra is a bit different [see Fig. ~\ref{fig:snr}(b)]. First of all, there are no peaks at the frequencies corresponding to integers of $50$ Hz, which supports the statement about the power supply circuit. At the same time, there is a peak near $1300$ Hz, which should be associated with the characteristics of the hair dryer. The spectrally averaged SNR in this case is $46$ dB. Hence, the lowest value of transmission spectra are limited by the SNR of the setup, leading to the distinction between the experimentally measured and numerically calculated results.

\begin{figure}[htbp!]
    \centering
    \includegraphics[width=0.9\linewidth]{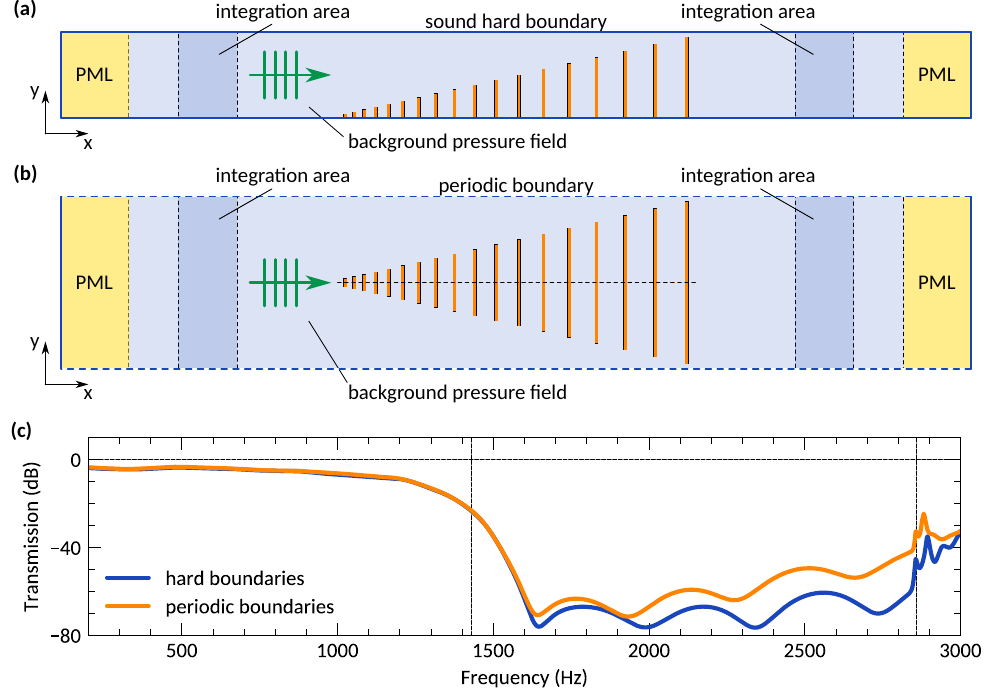}
    \caption{\textbf{Comparison of the models with different boundary conditions.} Schematic illustrations of (a) the numerical model of the waveguide with sound hard walls and (b) the model with the periodic boundaries. The parameters correspond to the case labelled as geometry II.
    Vertical dashed lines correspond to the frequencies at which the wavelength is equal to $w_{wg}/2$ and $w_{wg}$.}
    \label{fig:periodic_hard}
\end{figure}

To verify the fact that sound hard walls are equivalent to periodic boundary conditions, transmission spectra of the considered structure having the thickness of a single meta-atom is calculated for both conditions [see Figs.~\ref{fig:periodic_hard}(a) and  ~\ref{fig:periodic_hard}(b)]. The results in these cases are nearly the same, and only some slight variations within the stop-band can be observed  [see Fig.~\ref{fig:periodic_hard}(c)]. To be precise, the difference in the results is observed only when the semi-wavelength becomes smaller than the width of the waveguide with hard boundaries. Then, the discrepancy increases with the decrease of the wavelength, as it might be expected.

\section{Initial Design}
\begin{figure}[htbp!]
    \centering
    \includegraphics[width=0.9\linewidth]{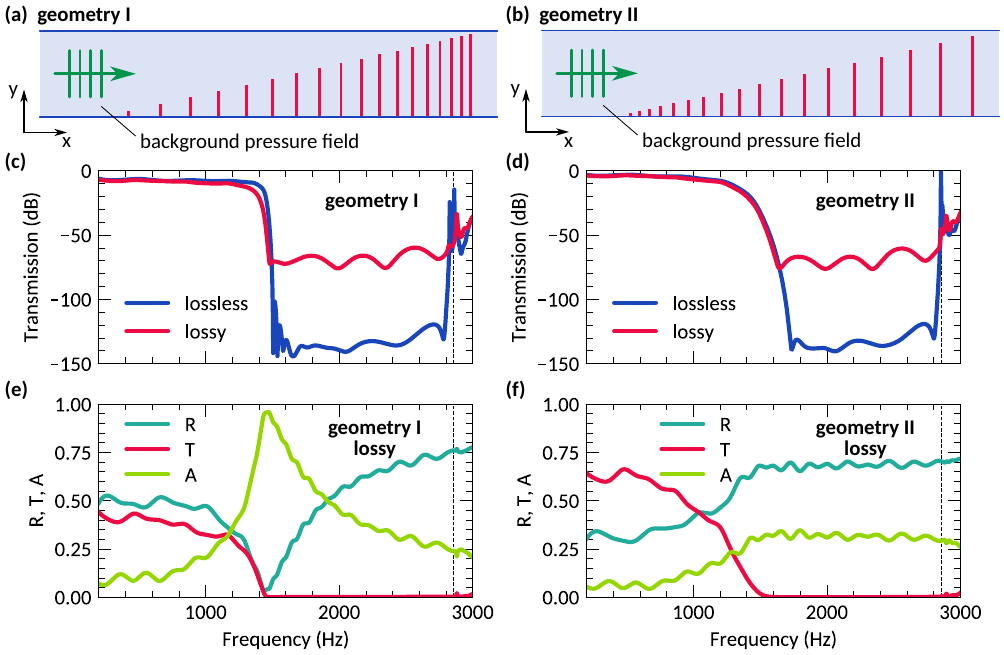}
    \caption{\textbf{Initial structures.} Schematic geometry of the considered structures characterized by the (a) increasing (geometry I) and (b) decreasing distances between the resonators (geometry II). (c), (d) Transmission spectra of the considered structures calculated for lossless case and for the case when thermoviscous losses are taken into account. (e), (f) Scattering and absorption coefficients of the considered structures for the case when thermoviscous losses are taken into account. Vertical dashed lines indicate the frequencies at which the wavelength of the incident field is equal to the width and double width of the waveguide.}
    \label{fig:initial_spectra}
\end{figure}

The design procedure is based on the circular duct acoustic black holes studied in Ref.~\cite{mousavi2022how}. Effectively, such axially symmetric structures can be considered as one-dimensional. To develop the two-dimensional counterpart, the parameters of the plates are considered to be the same. In particular, the coordinates of the plates along the $x$-axis are
\begin{equation}
    x_n = 5(n+1) + n(n-1)/2,
\end{equation}
where $n = 1, 2, \ldots, N$ is the sequential number of the plates, such that in total there are $N$ plates. The width of the plates are defined in two ways, namely
\begin{equation}
    w_n^{(I)} = W_{wg} - (115 x_n/255)/2,
\end{equation}
corresponding to the case when the distances between the plates decreases with the increase of their width [see Fig.~\ref{fig:initial_spectra}(a)], and
\begin{equation}
    w_n^{(II)} = (115 x_n/255)/2,
\end{equation}
such that the distances between the plates increase with the increase of their width [see Fig.~\ref{fig:initial_spectra}(b)]. Here $W_{wg}$ is the width of the waveguide in which the structures are placed, which is so far considered to be $60$ mm. All plates have the same thickness, which at this step is fixed at $d = 2$ mm. 

The calculated transmission coefficients of the considered structures are shown in Figs.~\ref{fig:initial_spectra}(c) and~\ref{fig:initial_spectra}(d). Both of the structures are characterized by the significant decrease of transmission in the spectra region from $1500$ to $3000$ Hz. At the same time, introduction of thermoviscous losses results in the increase of transmission, despite the fact that in the lossy case the absorption coefficient exceeds 0.3 in the spectral region of interest [see Figs.~\ref{fig:initial_spectra}(e) and~\ref{fig:initial_spectra}(f)]. Presumably, such behaviour might be associated with the suppression of resonances occurring in the cavities formed by the plates. 

\begin{figure}[htbp!]
    \centering
    \includegraphics[width=0.9\linewidth]{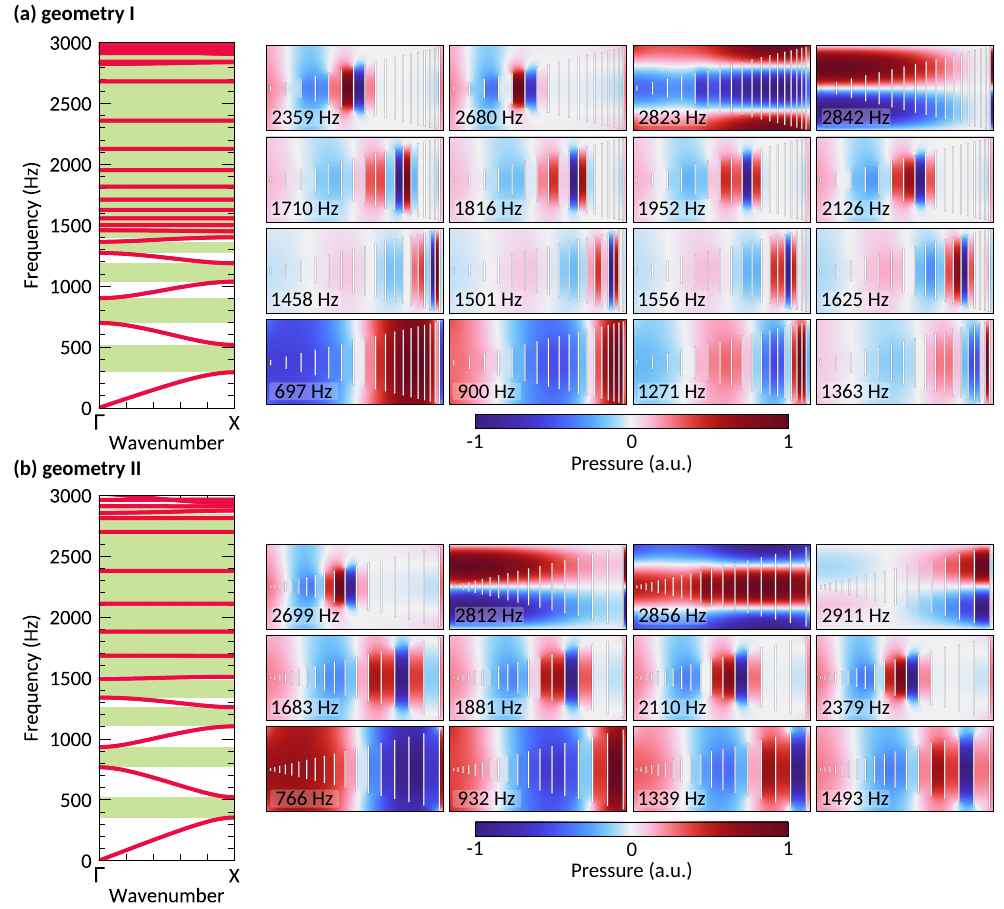}
    \caption{\textbf{Eigenmodes of the initial structures.} Band diagrams of the structures referred to as (a) geometry I and (b) geometry II in Fig.~\ref{fig:initial_spectra}. Shaded green areas indicate band-gaps. Field distributions are obtained for the $\Gamma$-point.}
    \label{fig:initial_eigenmodes}
\end{figure}

The last statement can be verified via consideration of the infinitely periodic structures with unit cells corresponding to the considered plate structures. In particular, the cells are characterized by the width $a_x = 250$ mm and the height $a_y = 120$ mm. As it might be expected, for both geometries, the spectral range from $1500$ to $3000$ Hz is characterized by the set of band-gaps (indicated by green shading) formed by flat modes [see Fig.~\ref{fig:initial_eigenmodes}]. Hence, the broad stop-bands of the finite-size structures should be associated mostly with the band-gaps of the equivalent infinitely periodic structures rather than with the thermoviscous losses. The field distributions of the associated eigenmodes are characterized by the field localized inside the area between plates, i.e. these are the modes associated with the resonances of the cavities formed by the plates, as it was assumed.

\begin{figure}[htbp!]
    \centering
    \includegraphics[width=0.9\linewidth]{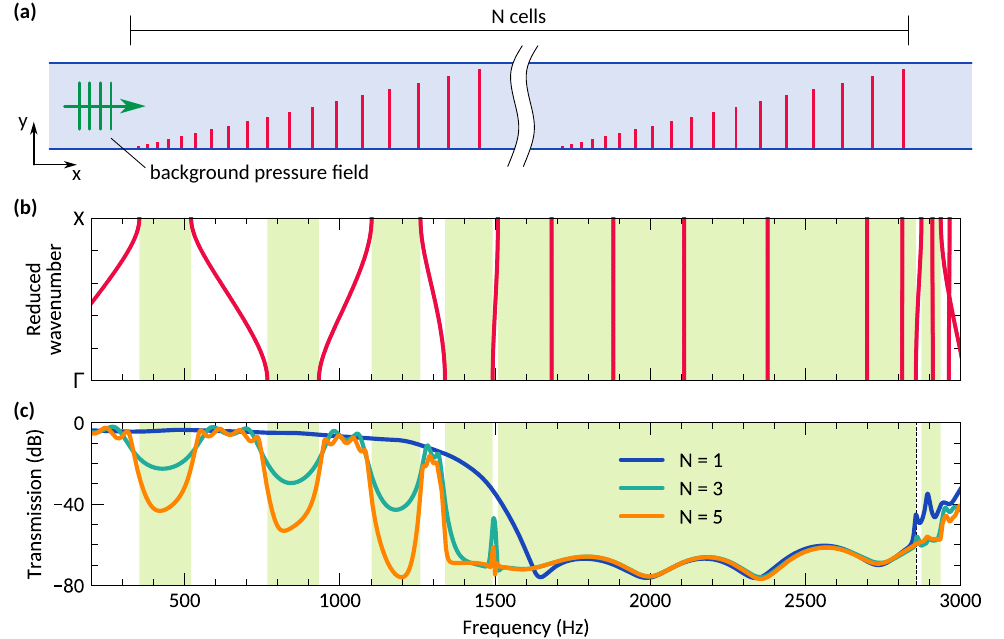}
    \caption{\textbf{Increase of the structures thickness.} (a) Schematic illustration of the considered finite-size having the thickness of $N$ unit cells. (b) Band structure of the infinitely periodic system. (c) Transmission spectra calculated for various thicknesses of the structure. Shaded green areas indicate band-gaps of the associated infinitely periodic structure.}
    \label{fig:initial_structure_Ncells}
\end{figure}

It should be also noted that both structures are also characterized by the additional band-gaps in the region below $1500$ Hz, though there is no decrease in the associated transmission spectra of the finite-size structure. Such discrepancy arises from the finite thickness of the structure, having a thickness of only one unit cell along the direction of wave propagation. The increase of the thickness results in the formation of additional stop-bands, as shown in Fig.~\ref{fig:initial_structure_Ncells}.

\begin{figure}[htbp!]
    \centering
    \includegraphics[width=0.9\linewidth]{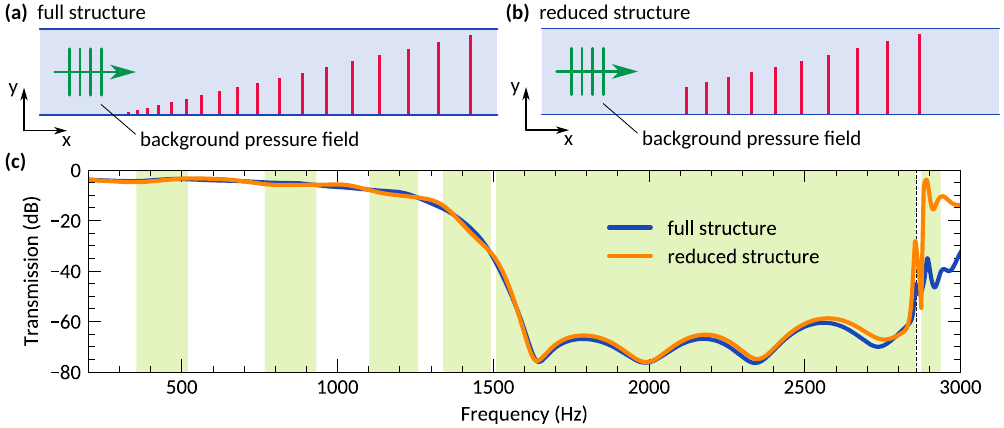}
    \caption{\textbf{Reduction of the structure.} Schematic illustration of the (a) full and (b) reduced structure, in which $8$ plates are removed. (c) Transmission spectra of the considered structures. Shaded green areas indicate band-gaps of the associated full structure.}
    \label{fig:initial_structure_reduced}
\end{figure}

Another conclusion which can be done on the basis of the eigenmodes analysis is that in the considered frequency range there are no modes associated with the resonances occuring in the cavities formed by the 8 smallest plates. Hence, these plates can be removed from the structure without affecting the transmission spectra, as confirmed by the results shown in Fig.~\ref{fig:initial_structure_reduced}.

\section{Equalization of the Cavities}
In order to get deeper insights into the origin of the band-gaps, it is reasonable to consider standalone resonators and their corresponding properties. Field distributions shown in Fig.~\ref{fig:initial_eigenmodes} indicate that at the frequencies corresponding to the flat bands the field is not localized inside a single cavity. Instead, the field distribution within a pair of neighboring cavities is close to anti-symmetric, which might be a manifestation of a strong coupling between two slightly detuned resonators. The detuning arises from the fact that the geometric parameters of the neighboring cavities are slightly different and hence the field distribution also can't be perfectly anti-symmetric. In such a case it might be expected that when the meta-atom is deformed in such a way that all cavities become equivalent, the coupling between them is maximized. At the same time, the strong local coupling within a unit cell is associated with the increase of band-gap sizes~\cite{hu2017acoustic}, and hence it might be expected that the band structure will be transformed accordingly.

\begin{figure}[htbp!]
    \centering
    \includegraphics[width=0.9\linewidth]{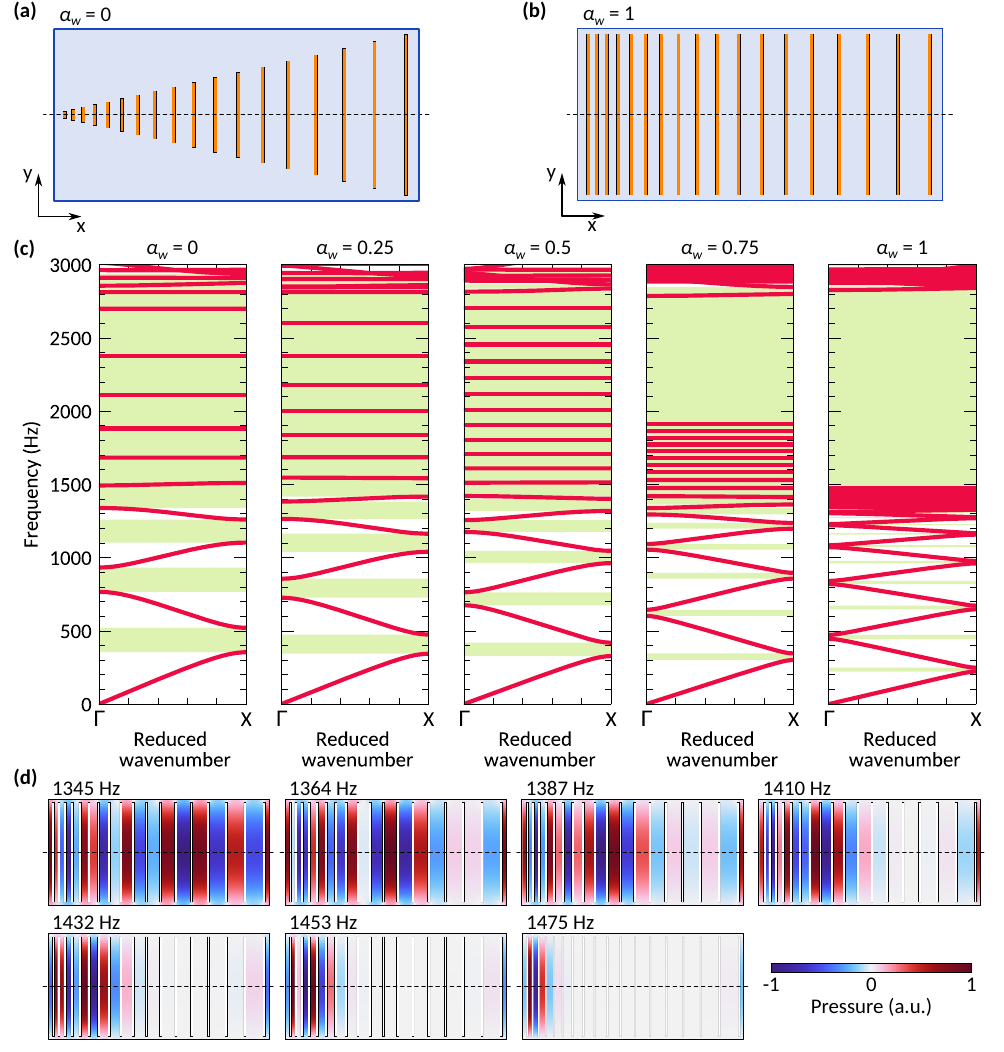}
    \caption{\textbf{Equalization of the plates.} Schematic illustration of (a) the initial structure and (b) the structure in which the width of the plates is equalized. The widths are modified via the parameter $\alpha_w$. (c) Evolution of the band structure with the increase of $\alpha_w$. Shaded green areas indicate the corresponding band-gaps. (d) Field distributions of the eigenmodes of structure with the equalized plates (i.e. $\alpha_w = 1$) obtained at $\Gamma$ point.}
    \label{fig:initial_tilt_eigenmodes}
\end{figure}

To verify the above statements, the meta-atom is transformed in such a way that the width of the plates changes from the gradually increasing to the equal ones, as shown in Figs.~\ref{fig:initial_tilt_eigenmodes}(a) and~\ref{fig:initial_tilt_eigenmodes}(b). Such a transformation can be introduced via additional parameter $\alpha_w$ defining the width of the plates as 
\begin{equation}
    w_n = (115 x_n/255)/2 + \alpha_w (w_{N} - w_n),
\end{equation}
where $w_N$ is the width of the largest plate. As previously, the size of the unit cell is still fixed at $a_x = 240$ mm and $a_y = 120$ mm. The increase of $\alpha_w$ results in the equalizing of the widths leading to the shifting of the flat bands to lower frequencies [see Fig.~\ref{fig:initial_tilt_eigenmodes}(c)]. Figure~\ref{fig:initial_tilt_eigenmodes}(d) indicates that the field in this case is localized within the larger number of neighbors as the detuning between the resonators decreases with the corresponding increase of the coupling. The drawback of the considered deformation is the closure of the low-frequency band-gaps so such meta-atoms should be less beneficial for the use in noise insulating structures.

\begin{figure}[htbp!]
    \centering
    \includegraphics[width=0.9\linewidth]{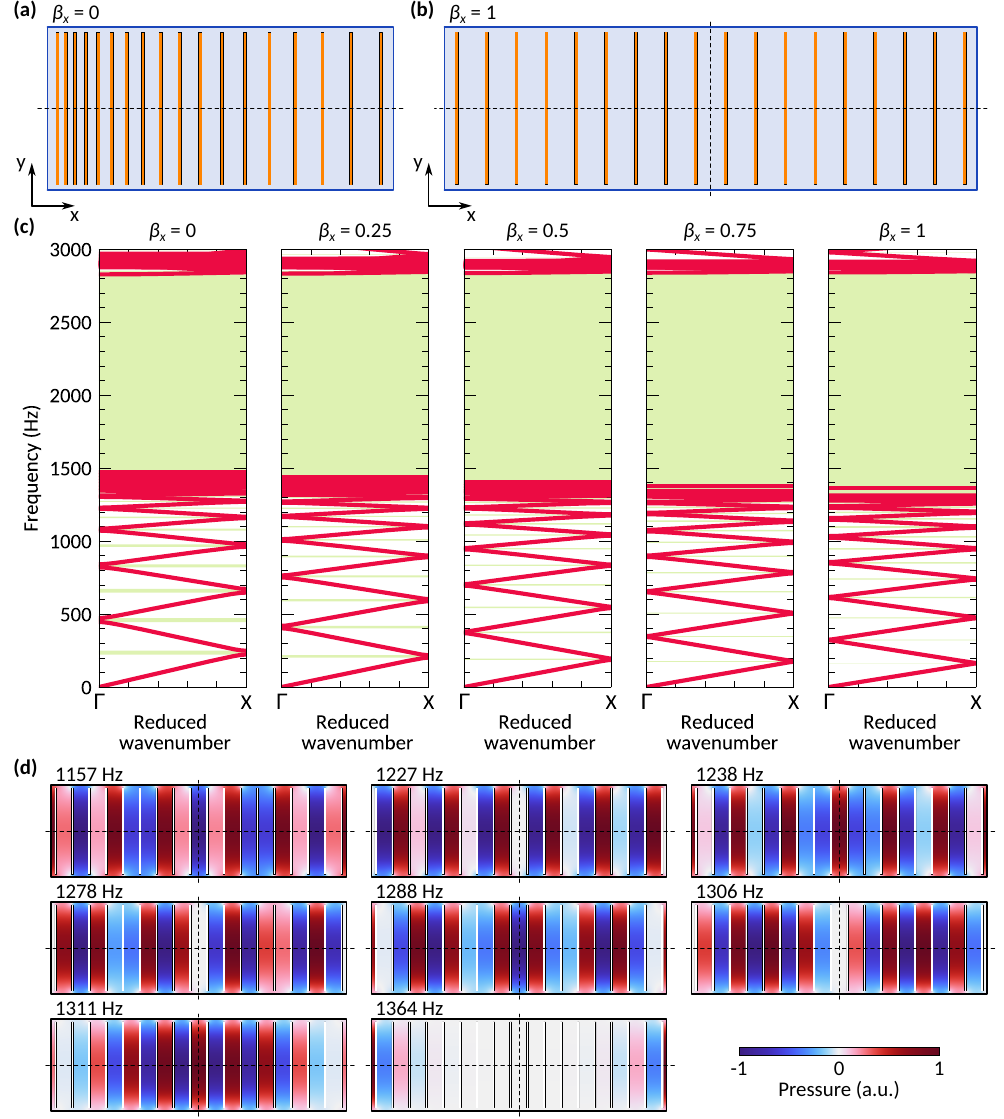}
    \caption{\textbf{Equalization of the cavities.} Schematic illustration of the structure with (a) gradually increasing distances between the plates and (b) the structure in which all distances between the plates are the same. The stretching of the structure along the $x$-axis is controlled by the $\beta_x$ parameter. The distance between the outer plates of the meta-atoms and the corresponding unit cell boundaries is fixed for all $\beta_x$. (c) Evolution of the band structure with the stretching of the meta-atom. Shaded green areas indicate the corresponding band-gaps. (d) Field distributions of the eigenmodes of the stretched structure (with $\beta_x = 1$) obtained at $\Gamma$-point.}
    \label{fig:initial_stretch_eigenmodes}
\end{figure}

However, the resonators are still not equivalent, which can be changed by the stretching of the meta-atom in a manner allowing to equalize the distances between the plates. For that the expression for the coordinates of the plates can be redefined as
\begin{equation}
	\tilde{x}_n = x_n + \beta_x \left[(n-1)(x_N - x_{N-1}) - (x_n - x_1) \right],
\end{equation}
where $\beta_x$ is the parameter controlling the stretching. The unit cell width $a_x$ is transformed in such a way that the distance between the boundaries of the unit cell and the first or the last plates is fixed at $5$ mm [see Figs. ~\ref{fig:initial_stretch_eigenmodes}(a) and ~\ref{fig:initial_stretch_eigenmodes}(b)].  According to Fig.~\ref{fig:initial_stretch_eigenmodes}(c), stretching of the meta-atom does not result in significant deformation of the band structure in the considered spectral range. However, there are notable changes in the corresponding field distributions [see Fig.~\ref{fig:initial_stretch_eigenmodes}(d)]. In particular, the unit cell becomes mirror-symmetric with respect to the vertical line passing through its center [see Fig.~\ref{fig:initial_stretch_eigenmodes}(b)]. As a result, all field distributions become either symmetric or anti-symmetric ones, while they are characterized by specific sub-patterns. For instance, the mode located at $1227$ Hz is characterized by a periodic set of anti-symmetric resonances localized in pairs of neighboring cavities, and the whole pattern is also anti-symmetric. The eigenmode at $1364$ Hz corresponds to the resonance of the cavity formed by the plates located in different unit cells, as the distance between them differs from the distances between the plates within the meta-atom. Other modes are also characterized by non-trivial patterns arising due to the coupling between the resonators and the associated mode splittings. Moreover, the meta-atom in this case is itself a finite-size period structure, which also increases the complexity of  the eigenmodes analysis. However, a rigorous investigation of the effects associated with the coupling of multiple resonators lies beyond the scope of the work. One important remark is that the modes below $1500$ Hz represent a set of inclined lines such that the slope decreases with the increase of frequency. In other words, the group velocity gradually changes until it reaches the value of zero, after which the band-gap opens. This might be considered as a direct analogy of the ABH effect, implying the decrease of the phase velocity with the propagation coordinate, and hence the considered stretched meta-atom might be referred to as a \textit{reciprocal ABH} in which vanishing group velocity might be associated with the strong coupling between the set of resonators.

\section{Coupling of the Resonances}
\begin{figure}[htbp!]
    \centering
    \includegraphics[width=0.9\linewidth]{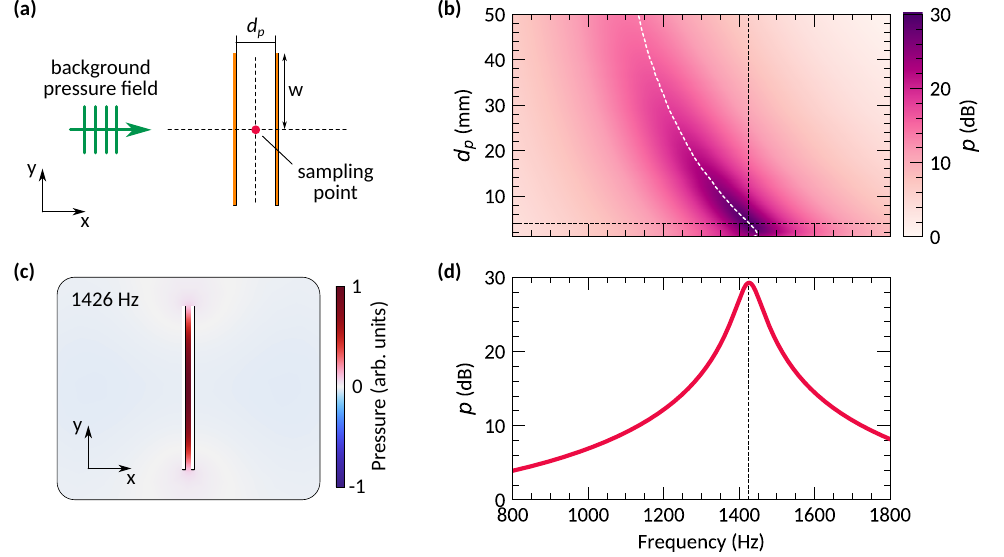}
    \caption{\textbf{Solitary resonator.} (a) Schematic illustration of a resonator formed by two plates placed in a free space. The plates are characterized by the semi-width $w$, while the distance between them is $d_p$. The resonator is excited by a plane wave and the pressure is calculated at the central point of the resonator. (b) Pressure spectra calculated inside the resonator. Dashed white line indicate the corresponding eigenmode. Black dashed lines indicate the maximum of the pressure occuring at $1426$~Hz when $d_p = 4$~mm. (c) Eigenmode field distribution for the case when $d_p = 4$ mm. (d) Pressure spectra for the case when $d_p = 4$~mm.}
    \label{fig:single_cavity}
\end{figure}

To support the claim about the strong coupling of the resonators, it is reasonable to consider cavities in free space. First of all, a solitary cavity is considered, such as the one shown in Fig.~\ref{fig:single_cavity}(a). The semi-width of the plates is fixed at $w = 56$ mm, which corresponds to the semi-width of the plates shown in Fig.~\ref{fig:initial_stretch_eigenmodes}. Such a simple system is characterized by a resonance occurring within the spectral range $1100$ - $1500$ Hz [see Fig.~\ref{fig:single_cavity}(b), where the white dashed line indicates the corresponding eigenmode]. The increase of the distance between the plates results in the shift of the resonant frequency towards the lower values, while the quality factor of the resonance also decreases. It might be expected that the maximal pressure enhancement inside the cavity is achieved at small distances between the plates. However, narrowing of the cavity also increases thermoviscous losses, which leads to the suppression of the resonance. The maximal pressure enhancement is observed at the frequency $1426$ Hz when the distance between the plates is $d_p = 4$ mm [see Figs.~\ref{fig:single_cavity}(b) and~\ref{fig:single_cavity}(d)]. In this case the eigenmode  [see Fig.~\ref{fig:single_cavity}(c)] is characterized by rather strong localization inside the cavity, such that the pressure is homogeneous around the center and gradually decreases near the open ends of the cavity.

\begin{figure}[htbp!]
    \centering
    \includegraphics[width=0.9\linewidth]{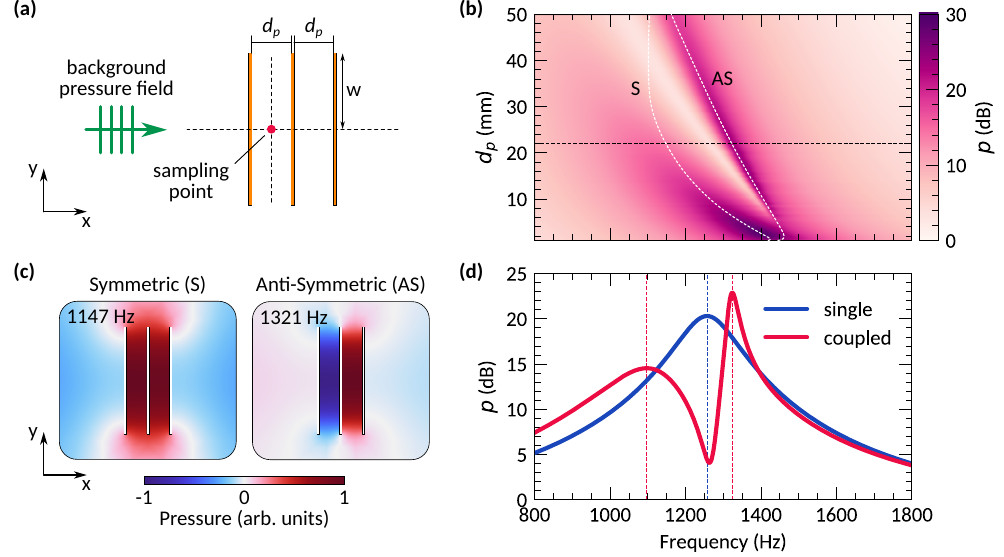}
    \caption{\textbf{Two coupled resonators.} (a) Schematic illustration of two coupled resonators formed by two plates placed in a free space. The plates are characterized by the semi-width $w$, while the distance between them is $d_p$. The resonators are excited by a plane wave and the pressure is calculated at the central point of the first resonator. (b) Pressure spectra calculated inside the first resonator. Dashed white line indicate the corresponding eigenmodes with the symmetric (S) and anti-symmetric (AS) field distributions. Black dashed line corresponds to $d_p = 22$~mm. (c) Eigenmode field distribution for the case when $d_p = 22$ mm. (d) Pressure spectra for the case when $d_p = 22$~mm in comparison with the pressure spectra of a solitary resonator with the same value of $d_p$. Vertical lines indicate the corresponding spectral peaks.}
    \label{fig:two_cavities}
\end{figure}

When the second cavity is introduced into the system [see Fig.~\ref{fig:two_cavities}(a)], it might be expected that the splitting of the resonance should occur. Indeed, the pressure spectra shown in Figs.~\ref{fig:two_cavities}(b) demonstrate the presence of two resonances, such that one of them is shifted towards lower frequencies, with respect to the resonance of a single resonator, and another one - towards higher frequencies [see ~\ref{fig:two_cavities}(d)].  Field distributions of the corresponding eigenmodes [see ~\ref{fig:two_cavities}(c)] indicate that one of the resonances is characterized by a symmetric field distribution, when both of the resonators are excited in-phase. The second eigenmode is characterized by anti-symmetric field distribution, such that the field is localized inside the cavities. When the distance between the plates decreases, the spectral width of the corresponding resonance also decreases until reaching zero [see ~\ref{fig:two_cavities}(b)]. This is a manifestation of a bound state in the continuum (BIC), which is a state characterized by the complete decoupling from the far field due to the interference of several leaky modes of the system~\cite{hsu2016bound, sadreev2021interference, koshelev2023bound}. It was shown, that in periodic structures BIC allow to increase the width of band-gaps~\cite{krasikova2023metahouse}, and in general can be utilized for the absorption enhancement and improvement of noise-insulating systems~\cite{krasikova2024acoustic}. 

\begin{figure}[htbp!]
    \centering
    \includegraphics[width=0.9\linewidth]{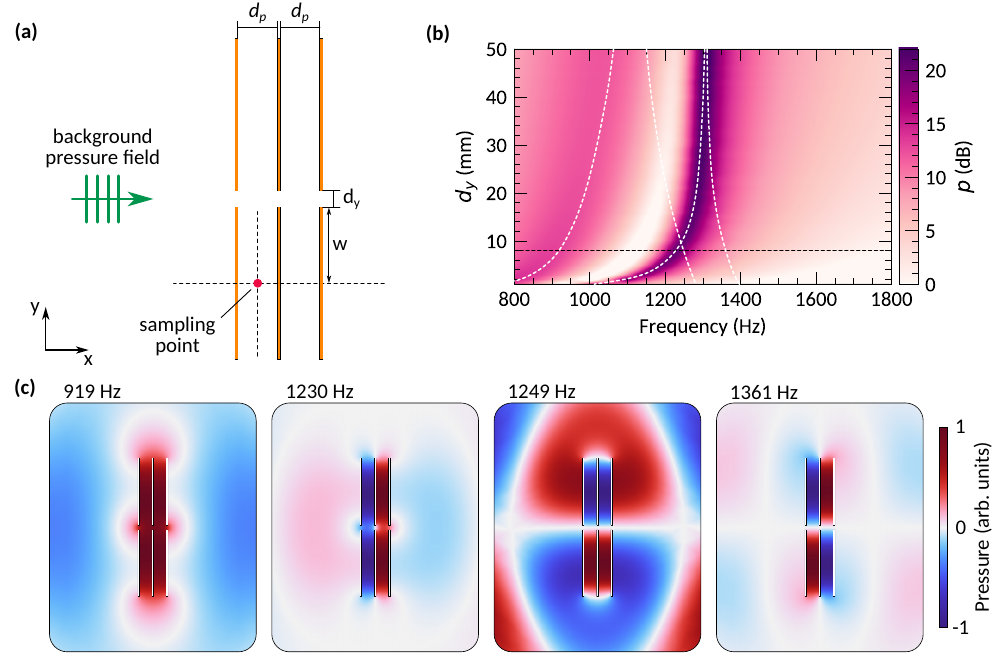}
    \caption{\textbf{Four coupled resonators.} (a) Schematic illustration of two pairs of coupled resonators separated by the distance $d_y$ along the $y$-axis. The plates are characterized by the semi-width $w$, while the distance between them is $d_p = 22$~mm. The resonators are excited by a plane wave and the pressure is calculated at the central point of one of the resonators. (b) Pressure spectra calculated inside one of the resonators. Dashed white line indicate the corresponding eigenmodes. Black dashed line corresponds to $d_y = 8$~mm. (c) Eigenmode field distribution for the case when $d_y = 8$ mm.}
    \label{fig:four_cavities}
\end{figure}

It should be noted that contrary to the one dimensional case, the periodicity of the two dimensional counterpart implies the possibility of interaction between the cavities along the $y$-axis (i.e. direction of periodicity). For instance, a system of four resonators might be considered, such that they are arranged in two pairs separated by the distance $d_y$ along the $y$-axis [see Fig.~\ref{fig:four_cavities}(a)]. According to the spectra shown in Fig.~\ref{fig:four_cavities}(b), the spectral position of the resonance is affected by the distance between the resonators, such that the smaller the distance, the lower is the position of the resonances. Such behavior can be explained via consideration of the eigenmodes, as the corresponding field distributions [see Fig.~\ref{fig:four_cavities}(c)] indicate that basically four resonators may act as the two elongated ones. Note that in this case there are four eigenmodes, but two of them are characterized by the reflection symmetry with respect to the horizontal line. Despite the presence of such modes, their excitation by a plane wave propagating along the $x$-axis is prohibited via symmetry considerations.

\begin{figure}[htbp!]
    \centering
    \includegraphics[width=0.9\linewidth]{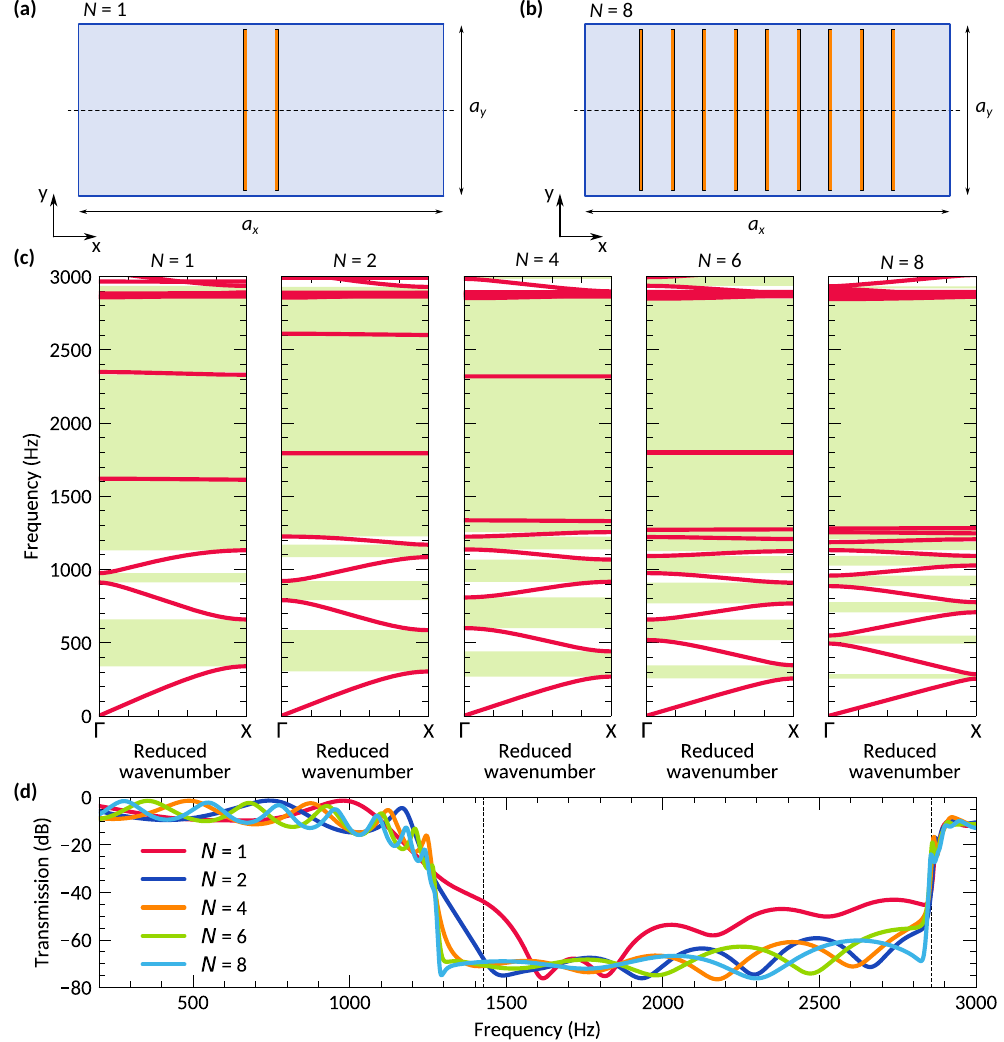}
    \caption{\textbf{Periodic structures made of equivalent coupled resonators.} Schematic illustration of a unit cell consisting of (a) one and (b) eight cavities. The width and the height of the unit cell is fixed at $a_x = 250$~mm and $a_y = 120$~mm. (c) Evolution of the band structure with the increase of the number of cavities. Shaded green areas indicate the corresponding band-gaps. (d) Transmission spectra of the finite-size structures characterized by the different number of cavities. Vertical dashed lines indicate the frequencies at which the wavelength of the incident field is equal to the width and double width of the waveguide.}
    \label{fig:periodic_N_cavities}
\end{figure}

Returning to the case of periodic structures, it hence should be expected that increase of the number of resonators should result in the increase of the number of modes and widening of band-gaps. In particular, the unit cell with the width $a_x = 250$ mm is considered, such that its height is still fixed at $a_y = 120$ mm. Within the unit cell there are $N$ cavities formed by $N + 1$ plates [see Fig.~\ref{fig:periodic_N_cavities}(a)]. It should be noted however that the actual number of the cavities is higher as practically two additional cavities are formed by the plates of neighboring unit cells. The parameters of the plates are still the same, with the semi-width $w = 56$ mm and the thickness $d = 2$ mm. In order to have a connection with the previously described results, the distance between the plates is fixed at $d_p = 22$ mm, which corresponds to the distance between the last two plates of the initial structure [see Fig.~\ref{fig:initial_spectra}(b)] or to the distance between the plates of the stretched structure [see Fig.~\ref{fig:initial_stretch_eigenmodes}].
As it might be expected, the increase of $N$ results in the formation of additional modes, such that the number of modes in the low-frequency region increases [see Fig.~\ref{fig:periodic_N_cavities}(c)]. The increase of the number of cavities also leads to the narrowing of low-frequency band-gaps, while the total width of band-gaps within the region $1000$ - $3000$ Hz remains nearly the same. From this point of view, the case of the structure with $N = 1$ allows to achieve the best result, when band-gaps cover almost the whole considered spectral range. When the corresponding finite-size structures are considered, their transmission spectra are characterized by the pronounced stop-bands, even for $N = 1$ [see Fig.~\ref{fig:periodic_N_cavities}(d)]. Though, the increase of the number of plates allows to broaden the stop-band and decrease the overall transmission. It also should be noted that the finite-size structures are not fully equivalent to the infinitely periodic ones as the unit cell contains additional cavities formed via plates of neighboring cells, which is not the case for finite-size structures.

\section{``Spruce'' Meta-Atom}
\begin{figure}[htbp!]
    \centering
    \includegraphics[width=0.9\linewidth]{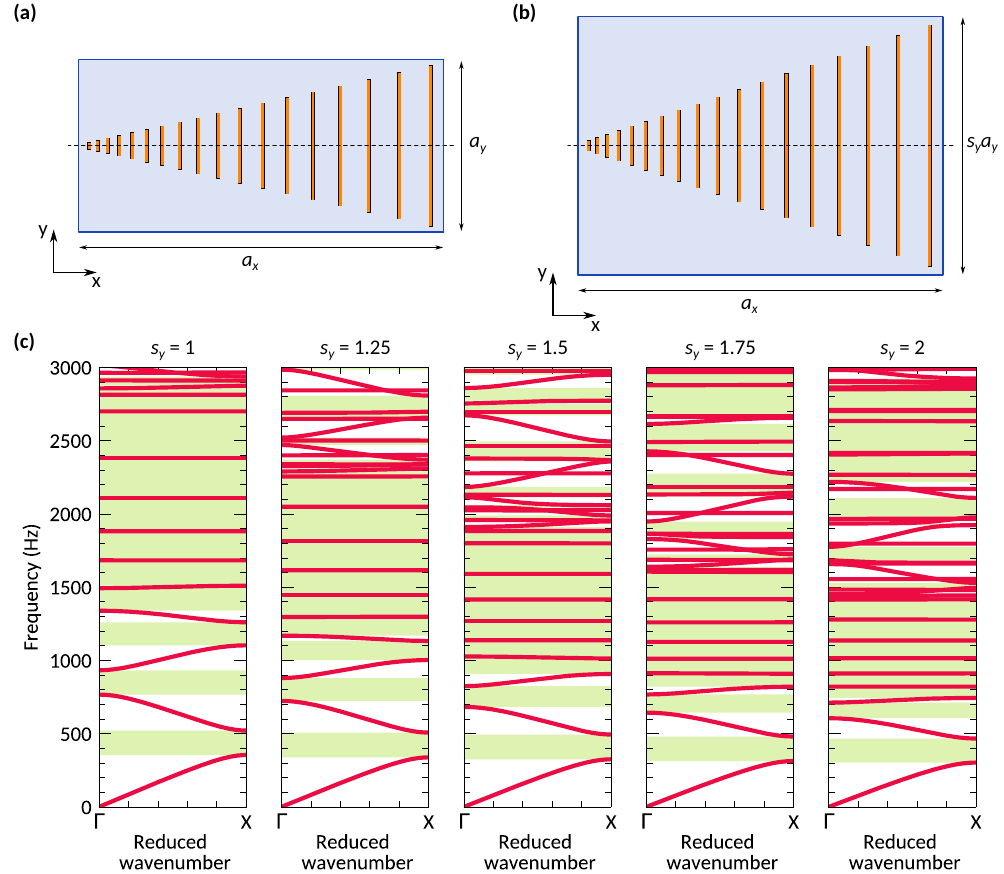}
    \caption{\textbf{Transformation of the initial meta-atom to the ``spruce''.} Schematic illustration of the (a) non-scaled and (b) scaled structure in which all geometric sizes along the $y$-axis are multiplied by the $s_y$ factor. (c) Evolution of the band structure with the increase of the scaling along the $y$-axis. Shaded green areas indicate the corresponding band-gaps.}
    \label{fig:initial2spruce}
\end{figure}

While even single resonator allows to achieve a broad stop-band, it is assumed that the results still might be improved. In particular, for the noise insulating applications it would be beneficial to shift the band-gaps towards the lower frequencies. This cane be done via simple scaling of the meta-atom, also taken into account the fact that the increase of the cavity size results in the shift of the resonant frequency towards the lower values (see Fig.~\ref{fig:two_cavities}). In particular, the scaling of the initial meta-atom along the $y$-axis is considered, such that all geometric parameters along this direction are multiplied by the factor $s_y$, as schematically shown in Figs.~\ref{fig:initial2spruce}(a) and~\ref{fig:initial2spruce}(b). According to the band diagrams [see Fig.~\ref{fig:initial2spruce}(c)], all modes shift towards the lower frequency region, and the spectral distance between them decreases. For convenience, the structure scaled by the factor $s_y = 2$ is labeled as "spruce", just by association with the schematic illustration of the corresponding tree. This structure is characterized by several band-gaps, the lowest of which occupies the range $300$ - $450$ Hz. Wide regions from $750$ to $1550$ Hz and from $2250$ to $2900$ Hz contain only the flat bands, so it might be expected that the transmission spectra of the corresponding finite-size structure will be minimized in these ranges, which is verified in the main text.

\bibliographystyle{unsrt}
\bibliography{bibliography}